\documentclass[twocolumn,final,showpacs,preprintnumbers,amsmath,amssymb]{revtex4}
 \usepackage[dvips]{graphicx}
 \usepackage{color}
\begin{document}

\preprint{}

\title{The three different  phases in the dynamics of chemical reaction
networks and their relationship to  cancer }
\author{David B. Saakian$^{1,2}$}
\author{Laurent Schwartz$^{3,4}$}
\email{saakian@yerphi.am}

\affiliation{$^1$Yerevan Physics Institute,2 Alikhanian Brothers
St., Yerevan 375036, Armenia}
 \affiliation{$^2$Institute of
Physics, Academia Sinica, Nankang, Taipei 11529, Taiwan}

\affiliation{$^3$ EcolePolytechnique, Laboratoire d' Informatique,
91128 Palaiseau France }
\affiliation{$^4$ Service d' oncologie,
Hopital Raymond Poincare, Garches }

\begin{abstract}

 We investigate the catalytic reactions model used in cell
modeling. The reaction kinetic is defined through the energies of
different species of molecules following random independent
distribution. The related statistical physics model has three phases
and these three phases emerged in the dynamics: fast dynamics phase,
slow dynamic phase and ultra-slow dynamic phase. The phenomenon we
found is a rather general, does not depend on the details of the
model. We assume as a hypothesis that the transition between these
phases (glassiness degrees) is related to cancer. The imbalance in
the rate of processes between key aspects of the cell (gene
regulation, protein-protein interaction, metabolical networks)
creates a change in the fine tuning between these key aspects,
affects the logics of the cell and initiates cancer. It is probable
that cancer is a change of phase resulting from increased and
deregulated metabolic reactions.

 \end{abstract}


 \maketitle

\section{Introduction}When observing the life on the ground of statistical physics of
complex systems we see a hierarchial level of organization and
modularity. How { can} we describe the "good, natural" relations
between different parts of complex system? In statistical physics we
describe different systems { through} different order parameter, and
in equilibrium different parts of the system have the same
temperature. For the stability of the system,  it is reasonable to
assume the similarity of order parameter at different hierarchy
levels and parts (modules) of the system. There is a nice similarity
with the no-arbitrage condition in financial markets \cite{de02}
where different stocks can fluctuate in equilibrium have only
identical order parameters,  defines as a ratio of driven and
diffusion motions. The Random Energy Model (REM)\cite{de80} and
related complexity (replica symmetry breaking) order parameters
\cite{sa05} work { starting} from the proteins \cite{gr00} till
quantum chromo-dynamics and strings, therefore we assume that the
cell organization  and cancer should not be exclusions, there is a
possibility that REM related ideas can work there. Using the known
results of REM, we will prove that the phase structure of the
chemical reaction network kinetics is related with the probabilistic
distributions of different chemicals in the steady state
distribution. Then we will speculate about the origin of the cancer
using similar complexity ideas.

One of the main ideas { for cell modeling} is to consider the
dynamical models with network structure \cite{ka69, ka06, kad09}.
One can realize this program studying the network of catalytic
reactions, identifying the phenotype  of the cell with the attractor
of the nonlinear system of differential equations.

 While constructing
the kinetic constants, it is { advantageous} to have a detailed
balance condition. The density of different chemical components
change according { to} the kinetic constants, and the latter are
defined through the energies of that components.
 We are following  \cite{ka09}. There are $M$ chemical
components. The transformation between chemicals $X_i$ and $X_j$ is
catalyzed by some component $x_c$, { so has} a rate $k_{i,j}x_cx_i$

Thus we can write a set of equations \cite{ka09}
\begin{eqnarray}
\label{e1} \frac{dx_i}{dt}=\sum_{j,c}
W(i,j,c)x_c((k_{j,i}x_i-k_{i,j}x_i)
\end{eqnarray}
where $W(i,j;c)=W(j,i;c)=1$ when there is a reaction and zero
otherwise. The kinetic coefficients are defined { through} the
energies of the chemical component $E_i$ and some inverse
temperature $\beta$ \cite{ka09}
$k_{i,j}=min\{1,\exp[\beta(E_i-E_j)]$.
 Thus  the kinetic
coefficients are defined { through} the energy landscape. The choice
of $k_{i,j}$  is quite reasonable for  catalytic reaction network:
there is some finite rate when the reaction goes to the low energy
configuration, { while there is} a small probability for the
reaction in inverse direction.

The Eq.(1) has a steady state solution
\begin{eqnarray}
\label{e2} x^s_i=\frac{\exp(-\beta E_i)}{\sum_j\exp[-\beta E_j] }
\end{eqnarray}
In \cite{ka09} { it was} considered a homogenous distribution of
energies in some interval $[0,\epsilon]$ { and} at some temperatures
glassy behavior { was found}. The authors claimed that the same is
the situation in case of normal distribution, while the phenomenon
lacks in case of log-normal distribution or the distribution with
the
tail.\\

\section{Results}

{\bf The statistical physics phases of the model.}

Consider a general distribution for energy levels
\begin{eqnarray}
\label{e3}
\rho(E_i)=\frac{1}{2\pi}\int_{-i\infty}^{i\infty}dh\exp[-hE+\ln
(M)\phi(h)]
\end{eqnarray}
where $\phi(h)$ is a some function. For the normal distribution we
have $\phi(\beta)=\beta^2/2$.

{ When considering} pure deterministic dynamics with $M$ degrees, we
can map our model to the statistical physics model REM \cite{de80}
with $\ln M$ degrees { assuming} large M and independence of energy
distribution.

In our model $E_i$ are the energies connected with different
chemical components of the model.

Nevertheless let us introduce a "partition function" Z,
$Z=\sum_{i=1}^Mx^s_i$. The statistical physics model with the
partition function Z has two phases. At small $\beta$ it is in the
paramagnetic phase (PM) with fast relaxation. When the following
equation has a real positive solution $\beta_c$ \cite{sa12}:
\begin{eqnarray}
\label{e4} (1+\phi(\beta_c))-\beta_c\phi'(\beta_c)
\end{eqnarray}
the second, spin-glass phase (SG) is possible in the model at
$\beta>\beta_c$. Two phases have different probability distributions
for the Z in the statics \cite{de89}. While formulated in spin-spin
interaction version, in SG phase there is a slow relaxation, see
also \cite{orland}. Thus we related some statistical physics with
the quenched disorder of the dynamic model (1).

How { can} we distinguish two phases? At different phases there are
different distributions functions for Z, { and} also different
expressions for the order parameter
$m=1-\frac{\sum_i(x^s_i)^2}{(\sum_ix^s_i)^2}$.

One has two expressions for the order parameter (connected with
replica symmetry breaking) \cite{gr84}
\begin{eqnarray}
\label{e5} m=1,\beta<\beta_c,\nonumber\\
 m=\frac{\beta_c}{\beta},{\bf \beta>\beta_c}
\end{eqnarray}
For the normal distribution case $\beta_c=\sqrt{2}$ \cite{de80}.
While there are $M$ degrees in our dynamic model instead of $\ln M$
spins in related REM, we observed the phase transition in the
dynamics, found first in \cite{ka09}. A similar model has been
solved in  \cite{orland}.

In the PM phase the order parameter $m(t)$ converges to the steady
state value as \cite{orland}
\begin{eqnarray}
\label{e6} |m(t)-1|\sim \exp[-a_1*t]
\end{eqnarray}
while in the SG phase \cite{orland}
\begin{eqnarray}
\label{e7} |m(t)-m|\sim \exp[-a_2*t^m]
\end{eqnarray}
where $a_1,a_2$ are some constants.

Let us now add a special energy level $E_0=-J_0\ln M$. When
$J_0>\beta_c$, third, ferromagnetic phase (FM), { is possible}
\cite{sa05}. We investigated the dynamics of the model in this case.
When the steady state concentration of the "ferromagnetic" chemical
is comparable with the total concentration of the system, the
reaction dynamics is becoming slower than even in the SG phase for
the initial homogenous distribution. The reason of ultra-slow
relaxation is the energy gap between the ferromagnetic level and
other levels. We performed a numeric for the case of normal
distribution (realistic for the biology of the cell
\cite{ka05},\cite{sal12}) with $M=100,K=10$ see Fig. 1, and observed
the phase transition phenomenon in the dynamics. The SG phase Fig 1a
is much more sensitive to the appearance of FM configuration, than
the PM phase, Fig 1b.

\begin{figure}
\large \unitlength=0.1in
\begin{picture}(42,12)
\put(-1.7,0){\includegraphics{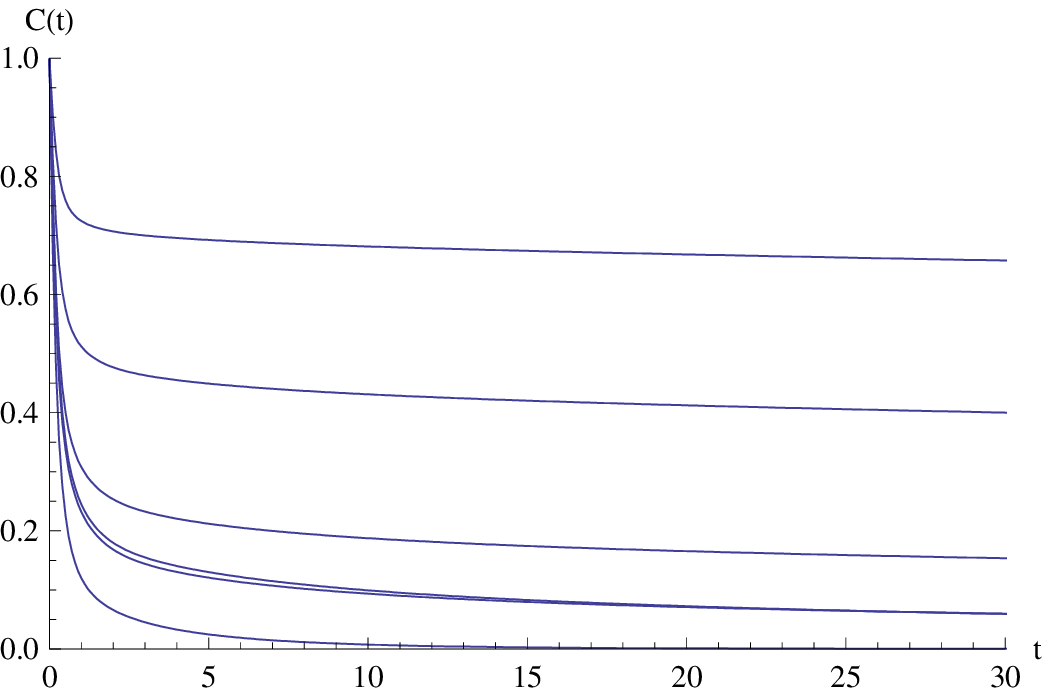}}
\put(16.5,0){\includegraphics{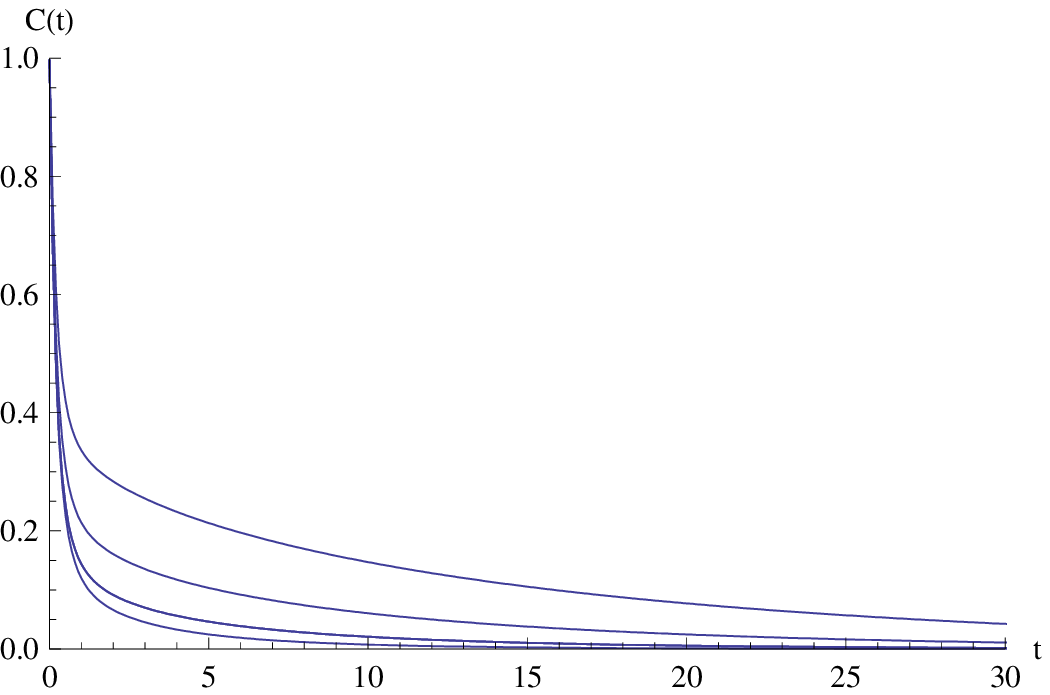}}
\put(7.7,5){\small{a.}} \put(26,5){\small{b.}}
\end{picture}
\caption{The $C(t)=\frac{\sum_i(x^c_i-x_i(0))(x^c_i-x_i(0))}{\sum
_i[(x^c_i-x_i(0)]^2}$ vs time. $N=100$. Every chemical is connected
with $K=10$ other chemicals.(a) The low line has no ferromagnetic
configuration, $\beta=0.8\beta_c$, the next line has $\beta=\beta_c
*1.4$, no ferromagnetic configurations. the higher lines correspond
to the $\beta=1.4\beta_c$ and ferromagnetic configurations with the
relative concentration $0.08,0.15,0.30,0.60$. (b)$\beta=\beta_c *0.8
$, there is no ferromagnetic configurations. the higher lines
correspond to the $\beta=1.4\beta_c$ and ferromagnetic
configurations with the relative concentration
$0.08,0.15,0.30,0.60$.}
\end{figure}
This model can be applied for the reactions in the cell. We assume
that the normal chemical kinetics of the cell corresponds to the SG
phase, contrary to PM or FM phases.

{\bf How general are these results?}

To obtain two phases we { assume} that: a. The nonlinear system of
equation has one steady state solution and b. randomness of steady
state concentrations $x^s_i$. The second condition is well {
established} experimentally for the concentration of chemicals in
the cell \cite{sal12}. The concentration of { chemicals} fluctuate
from  cell to cell. We used a concrete form of the nonlinear
dynamic. Let { us} consider another version of nonlinear dynamics,
i.e. $\frac{dx_i}{dt}=\sum_{j,c}
W(i,j,c)(x_c)^2(k_{j,i}x_i-k_{i,j}x_i)$. We performed numerics and
found the same qualitative picture: the statistical physics phase
for the quenched disorder defines the phase of the dynamic. If we
consider the relaxation from the initial configuration near the
steady state, then the results of different nonlinear attractor
models  certainly are similar; they are defined mainly by $k(i,j)$.

How { can} we apply these results to the cell-cell interaction, gene
expression problem?
 Ao formulated well the idea of adaptive landscape for
bio-networks \cite{ao09}. In the Ao's approach, while there is a
noise, transversal (magnetic like) field and friction, the system
has a given steady state defined by a potential landscape.{  This
potential landscape should be identified with the fitness. The
latter sometimes is directly connected with the ordinary free
energy.} This idea have been applied to the cancer analysis
\cite{ao08}. The robustness and plasticity aspects in the dynamics
of the cancer cells have been considered in \cite{ka09a} and
\cite{ga10}. The key point here is again  the concept of the
landscape again.
 Again fast and slow phases in the dynamics
are possible, for our derivation was important just the steady state
distribution of energy (fitness). { We considered SG-PM phase
transition at different temperatures. The same transition happens
when the steady state distributions are changed. This is an
important observation as the temperature is rather a constant in
biological systems.}

{\bf Connection between the phase in cell's chemical kinetics
and the type of fitness landscape.}

 According to our classification, the steady state could be
identified either with SG, FM, or PM ones. The other characteristic
of the model are the basin of attractions of these steady states. It
is an advantage to have large basins of attractions, but
simultaneously a slow relaxation dynamics.

A very interesting { point} is the connection of behavior of the
cell (basin of attraction + phase of quenched disorder) with the
evolution behavior. Here { it was} found that the dynamics of the
evolving population drastically depends on the character of the
fitness landscape function: the FM, or SG or their border case,
FM-SG \cite{sa09}. It is highly intriguing if the character of
fitness function of evolving cell will be the same as the character
of the chemical reaction kinetics. A similar ideas about the tight
relation between cell processes and evolution characteristics has
been suggested in a series of articles by K. Kaneko and co-workers,
\cite{ka09a}, where { they claim} that the large basin of
attraction, related with the given phenotype,  can be connected with
the mutational robustness.

In \cite{sa05} also speculations { are made} about resonance in
complex systems, assuming identical complexity parameters. { It has
been assumed that the possibility of parametric amplification of the
motion (resonance) and parametric fast attenuation (anti-resonance)
is one of the key features of normal living systems. Such a property
can exist for some parameters of dynamic system, without being
present for other values.  }

{\bf Speculations about cancer.}

How { can} we understand the cancer? For a recent review see
\cite{ha11}. Here we distinguish two aspects: the origin and the
dynamic behavior. The evolution dynamics aspect of cancer is well
know since \cite{no76}-\cite{gr12}. The key role of metabolism is
also well recognized \cite{la12}.
 Looking { for}
simple and general origin for the cancer, we assume the following
hypothesis. Since the start of the life, there is some fine tuning
between different key aspects of the life such as gene regulation,
metabolism, pH, cytoskeleton. It appears that these different
aspects of life are intertwined, and their interaction defines the
decision making, the logics of the cell \cite{nu08}.

The fine tuning of the cell should be understood as an existence of
proper statistical physics phases, identical complexity parameters,
oscillations, and this fine tuning is supposed to be increasing in
parallel with the evolution.  According to our hypothesis, in { the}
case of cancer: external aggression like chronic inflammation or
increased cellular metabolism caused by oncogene activation,
disrupts the degree of the fine tuning, change the reaction rates
and as a result, the decision making, the logics of the cell.

 Having less degree of fine tuning, the cell, in some sense,
returned back in evolution history, using old machinery. A candidate
of such a poor fine tuning, can be the change in the reaction rates
(glassiness), discussed in the previous section. The slow, glassy
like dynamics is typical for the cell to support the non-equilibrium
in the cell, an important aspect of the life \cite{bu05}. Another
side of phenomenon is connected with the memory processes, the
glassy (slow) phase { possess} such memory, while the fast phase {
does not}. The necessity of glassy dynamics is well recognized in
case of immune system \cite{de05}. We assume that the cancer cell {
has} another version of glassiness than the healthy cells. It is
intriguing that the osmotic pressure plays a crucial role for both
cancer \cite{lau11} and cytoskeleton glassiness \cite{bu05}. In case
of metabolic network the cancer cells choose a fast dynamics, and
perhaps the reaction network system is in the PM phase.

The fast reaction rate and less fine tuning are tightly related.
Normal cells can either be in anabolic or catabolic state. They use
oxygen and burn, for example, glucose. It results energy, water and
Carbonic gas. The alternative pathway is anabolism. They use energy
water and Carbonic gas to synthesize for example glucose. The cells
can synthesize other compounds like DNA , RNA, cholesterol.... In
normal cells the two pathways cannot be done at the same time. But
at different time of the day one cell can either be in anabolic or
in catabolic phase. In cancer cells, some enzymes (modules) are on
the anabolic mode, some others are on the catabolic mode. It results
a deregulation of the metabolism which could be seen as a  different
phase \cite{lau11}.

 Consider
the change of decision making mechanisms.  Probably this is the
start point of  cancer. The metabolic rate and decision making are
key features of living matter \cite{ha99},\cite{mu09}. They should
be considered both on a single cell level, and in a cell-cell  (
tissular) interaction \cite{he08}-\cite{ko10},{\bf  \cite{br12}}. We
should carefully analyze the metabolic network \cite{ba11} and
decision making in the healthy cell, and compare them with those in
the cancer cell. According to \cite{la12}, the cancer cell is less
dependent on the surrounding constraints than the healthy cell.{ An
intriguing possibility is the de-synchronization in decision making
between healthy cells and cancer cells, for example due to lack of
polarity in cancer cells \cite{oh11}. Fortunately the concept of
de-synchronization is already generalized for non-harmonic processes
\cite{am}.

We mentioned already the hypothesis of \cite{sa05} about a property
of living system to reveal both resonance and anti-resonance
features. In case of cancer  the second property (apostasis) is
lost.  In \cite{ka06} it was suggested to construct a rather
simplified models of cells so as to capture key features of the
cancer phenomenon, instead of looking into too complicated models.
In case we try to give a simplified phenomenological model of cancer
via some dynamical model, we should take care about the mentioned
feature of the model (a presence of resonance and anti-resonance). }
\\

\section{Discussion}

We checked the existence of different phases (degree of glassiness)
in the chemical reaction network dynamics: fast, slow and
ultra-slow, connected with the quenched disorder. While the first
two phases has been found numerically in \cite{ka09}, we determined
the conditions when they { emerge and} also found the third phase.
 The
phenomenon exists only for the special distributions of chemicals
with a nontrivial solution for Eq.(6), which related to the
distribution of proteins in the cells \cite{la12}.
 { Looking for} a simple and
complexity related key reason for the cancer, we suggested a
hypothesis that the healthy and cancer cells have different phases
in the chemical reaction network dynamics, connected with the glassy
properties. The glassy like properties of statistical physics models
are not  artifacts of the modeling, they have very deep and
universal meaning as complexity order parameters. One can apply the
spin-glass order parameters, statistical physics phases, to the
immune system \cite{de05}, proteins \cite{gr00}, cytoskeleton
\cite{bu05}, the reaction network (the current work result).
 In the { cells,} both normal
and malignant, there are many modules playing the role of logical
units. The cancer  cell has a poor fine tuning of different modules,
some of these modules change their meaning, affecting the decision
making aspects of the cell. We should perform a careful analysis of
different aspects in metabolic reactions \cite{wa56}, gene
regulatory network, protein-protein interaction network,
cytoskeleton dynamics to understand different choices in case of the
healthy cell and cancer, the difference in a decision making
schemes.  We should verify our hypothesis, that different aspects of
the cell (organism) should share the same complexity order
parameters. { We emphasize the idea of "energy" landscape in modern
modeling of cell dynamics, including population dynamics. We found
that the statistics of distribution of different degrees at peaks of
the landscape is as important as the landscape itself. The landscape
models like the ones \cite{ao08},\cite{ao09} should be completed
also by a description of statistics of different components at
peaks. We suggest to try experimentally measure both the
distribution of different chemicals in healthy and cancer cell, and
also compare the rate of (metabolic) reactions, trying in this way
derive quantitative criteria to identify the transition between SG
(healthy, slow) and PM (fast) phases. Besides identifying different
statistical physics phases of chemical reaction networks, we suggest
to focus on decision making synchronization by cells and the
existence of anti-resonance property.}

It is possible that the investigation of the advanced evolution
models, incorporating some decision making aspects, will improve our
understanding of both the cancer and the origin of life.  Only
genetical aspects or metabolism are not sufficient to describe the
origin of life \cite{sz10}.

 DBS thank
the DARPA prophecy project,  NSC 101-2923-M-001-003-MY3 and NCTS
(North) in Taiwan, for the support,  K. Kaneko, H. Q. Chen, T. J.
Kobayashi
 for discussions.

\end{document}